\documentstyle[psfig,12pt,fullpage]{article}
\newcommand{\be}{\begin{eqnarray}}
\newcommand{\ee}{\end{eqnarray}}
\newcommand{\ba}{\begin{array}}
\newcommand{\ea}{\end{array}}
\newcommand{\nn}{\nonumber}
\newcommand{\ra}{\rightarrow}

\def\s{\scriptstyle}
\def\ss{\scriptscriptstyle}

 \def\d{\dagger}
\def\gt{\tilde\Gamma}
\def\f{\phi} \def\p{\psi} \def\k{\kappa}
\def\o{\over} \def\i{\infty} \def\a{\alpha}
\def\zf{Z_{\s\phi}}
\def\zp{Z_{\s\psi}}
\def\zfft{Z_{b,t}^{\ss \f\f\f\f}}
\def\zfpt{Z_{b,t}^{\ss \f\f\p\p}}
\def\zpft{Z_{b,t}^{\ss \p\p\f\f}}
\def\zppt{Z_{b,t}^{\ss \p\p\p\p}}

\def\bff{B^{\ss \f\f\f\f}}
\def\bfp{B^{\ss \f\f\p\p}}
\def\bpf{B^{\ss \p\p\f\f}}
\def\bpp{B^{\ss \p\p\p\p}}

\begin{document}
\setcounter{page}{0}
\thispagestyle{empty}
\begin{flushright}
ICN-UNAM-97-03\\
July 11, 1997
\end{flushright}
\vskip 0.5 truein
\begin{center}
{\LARGE
Regge Behaviour from an Environmentally Friendly Renormalization
Group\footnote{This work was
supported by Conacyt grant 3298P--E9608.}}
\vskip 0.4truein
{\large
\bf  C.R.\ Stephens\footnote{e--mail: stephens@nuclecu.unam.mx},
A. Weber\footnote{Supported by fellowships of the DAAD and the Mexican
Government; e--mail: axel@nuclecu.unam.mx}
J.C. L\'opez Vieyra\footnote{e--mail: vieyra@nuclecu.unam.mx},
and P.O. Hess\footnote{e--mail: hess@nuclecu.unam.mx}}\\
\vskip 0.25truein
{Instituto de Ciencias Nucleares, U.N.A.M., }\\
{A. Postal 70-543, 04510 Mexico D.F., Mexico.}\\

\end{center}

\vskip 1.3truein
\begin{sloppypar}
{\bf Abstract:} The asymptotic behaviour of cubic field theories is
investigated in the Regge
lim\-it using the techniques of environmentally friendly renormalization,
environmentally friend\-ly in the present context meaning asymmetric in its
momentum dependence. In particular we consider the crossover between
large and small energies at fixed momentum transfer for a model scalar theory
of the type
$\phi^2\psi$. The asymptotic forms of the crossover scaling functions are
exhibited for all two particle scattering processes in this channel to one loop
in a renormalization group improved perturbation theory.
\end{sloppypar}

\vfill\eject

One of the most active areas of interest in QCD both experimentally
and theoretically is the limit $Q^2\gg\Lambda^2_{\s QCD}$, $x\ll 1$ where
$Q^2$ and $x$ are the Bjorken scaling variables. This corresponds to the
diffractive, or Regge limit, where standard renormalization group (RG)
improved perturbation theory breaks down. Various methods have been
proposed and used to tackle this problem with varying degrees of success.
One of the most common is the summation of leading logs \cite{lipatov} via
an inspection of the perturbation expansion, a technique which has a long
history
(see for instance \cite{eden} and references therein). Besides an intrinsic
degree of arbitrariness, summing sets of logs can be quite difficult
combinatorially
in many cases.

The RG has proved to be an extremely useful non-perturbative tool,
especially in the context of non-abelian gauge theories where the ultraviolet
fixed point can be accessed via very simple renormalization procedures such as
minimal subtraction and has had great success when applied to deep-inelastic
scattering. In the Regge limit however, the RG has been conspicuous by its
absence, except for the beautiful application to Reggeonic field theory
\cite{migdal}.
The principal reason is its association with ultraviolet divergences. In the
Regge
limit the breakdown of perturbation theory has nothing to do with the latter,
however, terms such as $\ln s$ or $\ln t$ in the limits $s$, $t\ra\i$ do lead
to
divergences. These divergences in contradistinction to short distance behaviour
are very asymmetric. The reason they appear can be traced to the nature of
the effective degrees of freedom in the problem. For small $s$ and $t$ they
are four dimensional, whereas in the Regge limit,
as is well known, there is a ``kinematic'' dimensional reduction to two
dimensions
owing to the extreme anisotropy between the longitudinal and transverse
sectors.
Such crossovers between effective degrees of freedom of one type and another,
qualitatively completely different, are ubiquitous in physics. Indeed the
crossover
between asymptotic freedom and confinement offers a perfect paradigm.

To describe systematically such crossovers using RG methods one
requires a RG that can interpolate between different effective degrees of
freedom
as a function of ``scale'', where scale could mean temperature, momentum, size
etc.
Such an RG, that can be applied to a myriad of other crossover situations, has
been developed
under the name of ``environmentally friendly'' renormalization \cite{envfri} in
recognition of
the fact that a crossover very often can be thought of as taking place due to
the
effect of some ``environmental'' parameter, such as temperature. Using these
methods it has been possible, for instance, to access the dimensional crossover
in finite
temperature field theory \cite{fintemp} between an effective four dimensional
theory at low temperatures to an effectively three dimensional theory near a
second
or weakly first order phase transition. The purpose of this letter is to
present an
environmentally friendly renormalization that can access the non-perturbative
crossover
to Regge behaviour in the context of a scalar cubic model theory.  Applications
to QCD will
be considered in other publications.

First we will present the basic renormalization scheme
we will use to treat the $t$-channel crossover between ``small'' $t\sim s$
and large $t\gg s$ for fixed $s$, where $s$ is a momentum variable in the
physical region. The results
we present can also be applied to the $s$-channel crossover beween small
and large $s$. The theory we will consider is a simple ``mesonic'' cubic theory
where
we can ignore the effects of spin. We will consider an interaction of the
form $(g_B/2)\phi^2\psi$ where the $\phi$ and $\psi$ fields have bare masses
$m_B$ and $M_B$ respectively. Results for the cases $\phi^3$,
$\phi^{\d}\phi\psi$ and a Wick-Cutkosky type model $\sum_{i=1}^2
\phi^{\d}_i\phi_i\psi$ will follow quite simply from the case treated.

The function of interest here will be the connected four point function
$G^{\s ijkl}$ on shell. We use the superscript notation $ijkl$ to
denote the external legs, i.e.\ $i,j,k,l$ can take the values $\phi$ or $\psi$.
The last two indices refer to the incoming and the first two to the
outgoing particles in the $s$ channel as shown in figure 1.
To simplify matters we may remove the external legs to define
\be
\gt_B^{\s ijkl}(p_1,p_2,p_3,p_4)=\Gamma^{\s ii}_B(p_1)\Gamma^{\s jj}_B(p_2)
\Gamma^{\s kk}_B(p_3)\Gamma^{\s ll}_B(p_4)G_B^{\s ijkl}(p_1,p_2,p_3,p_4) \:.
\ee
The relation to fully one-particle irreducible vertex functions is via
\begin{eqnarray}
\gt^{\s ijkl}_B(s,t,u)&=&\Gamma_B^{\s ijkl}(s,t,u)+\Gamma_B^{\s ijm}(s)G^{\s
mm}_B(s)
\Gamma_B^{\s mkl}(s) \nn \\
&& {}+\Gamma_B^{\s ikm}(t)G^{\s mm}_B(t)\Gamma_B^{\s mjl}(t)+
\Gamma_B^{\s ilm}(u)G^{\s mm}_B(u)\Gamma_B^{\s mjk}(u)
\end{eqnarray}
for the external momenta on mass shell. One may further decompose the
irreducible four point
function in the following way (see below):
\be
\Gamma^{\s ijkl}_B(s,t,u)=A_{B,t}^{\s ijkl}(s,t)+A_{B,u}^{\s ijkl}(s,u)+
A_{B,s}^{\s ijkl}(t,u) \:.
\ee

Now we come to the question of renormalization of these functions. We
will consider $d=4$. As far as ultraviolet divergences are concerned the only
badly
behaved diagrams are those that contain radiative corrections to the masses,
however for $t\gg s$, at $n$-loop order, there exist diagrams which
give corrections of $O((\ln t)^n)$.
In the limit $t\ra\i$ we thus have new divergences which have nothing to do
with the ultraviolet. As mentioned in the introduction they are a
symptom of the fact that the correct
effective degrees of freedom of this model for asymptotically large $t$ are
completely different to those of the small $t$ region. This is due to
the fact that the transverse degrees of freedom become strongly coupled leading
to an effective ``dimensional reduction'', or perhaps better to say
``factorization'',
of the loop graphs into the form $g(t)K(s)$ where $g(t)$, associated with the
longitudinal
direction is effectively two-dimensional, whilst $K(s)$, associated with the
transverse
dimensions, is $(d-2)$-dimensional. Indeed, the dimensional crossover shares
several
features in common with that of say $\lambda\phi^4$ theory on $S^1\times
S^1\times R^{d-2}$,
where the four point coupling is to one loop
\be
\Gamma_B^{(4)}=\lambda_B-{3\o 2}{\lambda_B^2\o L^2}\sum_{n_1,n_2}
\int {d^{d-2}k\o \left(k^2+m^2+{4\pi^2(n_1^2+n_2^2) \o L^2}\right)^2} \:.
\ee
In the limit $Lm\ll 1$ one can neglect other than the zero eigenmodes and so
one has
a factorization into a $(d-2)$-dimensional loop integral and an $L$ dependent
effective
coupling constant $g_B=\lambda_B/L^2$. In this case for $Lm\gg 1$ the effective
degrees
of freedom of the system are $d$-dimensional whilst for $Lm\ll 1$ they are
$(d-2)$-dimensional.
The effective degrees of freedom being environment dependent, i.e. $L$
dependent,
require an $L$ dependent renormalization that is capable of capturing both a
$d$ and
$(d-2)$-dimensional fixed point. Such a renormalization it should be emphasized
is highly
anisotropic in that the relevant counterterms, unlike for instance minimal
subtraction type
counterterms, are very sensitive to the asymmetry between the finite and
infinite directions.
In the case at hand the effective degrees of freedom are strongly $t$ dependent
hence
an asymmetric renormalization that can take into account the anisotropy between
the
longitudinal and transverse directions is required. Clearly any renormalization
scheme
symmetric in the momenta such as used for the running coupling in QCD will be
totally
inadequate.

The vertex correction in the $t$ channel (see eq.\ (2)) varies like
$(g_B^2/t)^{n}(\ln t)^{2n}$ in four dimensions
at $n$ loop order. Thus there is no Sudakov suppression here and the bare
vertex
will give a good approximation. A similar argument holds true in the $u$
channel where
$u\ra -t$ asymptotically. In the $s$ channel the corrections do not go to zero,
however,
they are small in the sense that they are perturbatively controllable. For mass
renormalization of the $\psi$ field one may remove the ultraviolet divergence
via a mass shell renormalization, similarly for the mass divergence of the
$\phi$ field.
In the $t$ channel the resulting $n$-loop contribution to $G^{\s ii}_B$
is $\sim (g_B^2/t)^{n}((\ln t)^n/t)$. Thus the momentum dependent
corrections after an ultraviolet subtraction go to zero as $t\ra \i$. The same
holds
true in the $u$ channel, whilst in the $s$ channel the corrrections are
non-zero but once
again are
perturbatively controllable. The upshot of all this is that the crossover to
dimensionally
reduced behaviour in the large $t$ limit is controlled by ladder-type diagrams.
This is interesting in that renormalization of the theory cannot now be
achieved by
a reparametrization of the original parameters of the theory, $g_B$, $m_B$ and
$M_B$.

One concludes that a renormalization of the effective four point interaction
itself is required.  There is nothing particularly strange about this. One can
easily convince oneself of its naturalness by considering
the limit $M\ra\i$, $m\ra 0$ after renormalizing the theory as a cubic theory.
Physically
one knows that an effective $d$-dimensional $\lambda\phi^4$
theory has to emerge. However, renormalization of $m_B$ and $g_B$ is not
sufficient to remove the characteristic infrared divergences for $d\leq4$.
A subsequent renormalization of the
four point vertex itself is required in order to arrive at the usual $\phi^4$
renormalization
constants. This renormalization is associated only with the one-particle
irreducible part, given at one loop by the ``box'' diagrams. Of course,
this limit is completely opposite to the one we are considering here, i.e. the
Regge limit. It simply illustrates that there exists another kinematical regime
wherein one sees that renormalization of the parameters of the model
is not sufficient to render the theory perturbatively tractable.

There are two ways we can proceed now, by renormalizing $\gt_B^{\s ijkl}$
directly or via
a renormalization of $\Gamma_B^{\s ijkl}$. Here we will consider the
renormalization
at the level of the functions $A$ above in (3), or more precisely for the
functions
$B$ defined via
\be
B^{\s ijkl}_{B,t}(s,t)=A^{\s ijkl}_{B,t}(s,t) +
\Gamma_B^{\s ikm}(t)G^{\s mm}_B(t)\Gamma_B^{\s mjl}(t) \:,
\ee
\be
B^{\s ijkl}_{B,u}(s,u)=A^{\s ijkl}_{B,u}(s,u) +
\Gamma_B^{\s ilm}(u)G^{\s mm}_B(u)\Gamma_B^{\s mjk}(u) \:,
\ee
\be
B^{\s ijkl}_{B,s}(t,u)=A^{\s ijkl}_{B,s}(t,u) +
\Gamma_B^{\s ijm}(s)G^{\s mm}_B(s)\Gamma_B^{\s mkl}(s) \:.
\ee
A diagram contributing
to $\Gamma_B$ is by definition associated with $A_{B,t}$ if it
contains powers of $\ln t$ in the large $t$ limit, with $A_{B,u}$ if it
contains
powers of $\ln (-t)$, and with $A_{B,s}$ otherwise.

We now wish to define renormalized functions $B^{\s ijkl}$, restricting
attention to
$B^{\s ijkl}_t$ for the moment. Due to
mixing of the effective cuartic interactions with different $ijkl$ a pure
multiplicative renormalization is not sufficient, one must introduce a matrix
renormalization via
\be
B^{\s ijkl}_t(s,t,g(\k),m(\k),M(\k),\k)=\sum_{m,n}Z^{\s ijmn}_{b,t}(\k)
B^{\s mnkl}_{B,t}(s,t,g_B,m_B,M_B,\Lambda) \:.
\ee
We have introduced a cutoff in the bare functions to regularize any UV
sub-divergences. The renormalized parameters $m$, $M$ and $g$ may or may not
depend on $\k$ according to the specific renormalization procedure that we use.
In the case at hand in four dimensions we may renormalize $m$ and $M$ on the
mass shell without running into any problems in the limit $t\ra\i$. Similarly,
as
mentioned above no explicit $g_B$ renormalization is needed though one can
certainly implement one if it turns out to be convenient. One may also
introduce wavefunction renormalizations $\zf(\k)$ and $\zp(\k)$ for the fields
$\phi$ and $\psi$ respectively, but
once again in four dimensions such renormalizations are not necessary. Note
that we have suppressed in the notation dependence in the $Z$ factors on
parameters
other than $\k$, the specific functional dependence depending on the actual
specific
renormalization scheme we choose.

The functions $B^{\s ijkl}_t$ satisfy renormalization group equations
\be
\k{dB^{\s ijkl}_t(s,t,\k)\o d\k}=\sum_{m,n}\gamma^{\s ijmn}_{b,t}(\k)B^{\s
mnkl}_t(s,t,\k)
\:,
\ee
where $\gamma_{b,t}(\k)=(d Z_{b,t}(\k)/d\ln\k)\cdot Z^{-1}_{b,t}$ summarizes
the anomalous scaling behaviour of the functions with respect to $\k$.

To produce explicit results we introduce a specific renormalization
procedure to fix the values of the renormalization constants. We choose
\be
\Gamma^{\ss\p\p}(p^2=-M^2)=0
\ee
\be
\Gamma^{\ss\f\f}(p^2=-m^2)=0
\ee
\be
\left(\matrix{\bff_t(s,t=\k,\k)&\bfp_t(s,t=\k,\k)\cr
                     \bpf_t(s,t=\k,\k)&\bpp_t(s,t=\k,\k)\cr}\right)
=\left(\matrix{{g_B^2\o\k+M^2}&{g_B^2\o\k+m^2}\cr
                                    {g_B^2\o\k+m^2}&0\cr}\right)
\ee
%
The remaining effective interactions of the type $B^{\ss\f\p\f\p}_t$ etc.\ do
not
mix with the above and will not be considered in the following.
Note that the normalization conditions (10-12)
are explicitly $s$ and $t$ (via $\k$) dependent. In this sense they are an
environmentally friendly set of normalization conditions. The motivation behind
them is
that we are seeking a RG map to a region of parameter space
where a perturbative calculation is possible. The ``mean field'' regime will
always
offer such a region if the map to it can be perturbatively constructed. That
this map
can be used to give perturbative control over the Green's functions for
arbitrary values
of $t$ will of course put a constraint on the value of the arbitrary
renormalization scale
$\k$ we choose.

With the normalization conditions (12) in the large $t$ limit one finds to one
loop
\be
\left(\matrix{\zfft(s,\k)&\zfpt(s,\k)\cr
                     \zpft(s,\k)&\zppt(s,\k)\cr}\right)
=\left(\matrix{1-g_B^2K_m(s)\ln\k &-g_B^2K_M(s)\ln\k \cr
                   -g_B^2K_m(s)\ln\k &1\cr}\right)
\ee
where the functions $K_m$ and $K_M$ are given by two-dimensional one loop
diagrams:
\be
K_{\mu}(s)={1\o16\pi^2}\int_0^1{d\beta\o \beta(1-\beta)s+\mu^2} \:.
\ee
The anomalous dimension matrix is
\be
\gamma_{b,t} =
\left(\matrix{-g_B^2K_m(s)&-g_B^2K_M(s)\cr
                    -g_B^2K_m(s)&0\cr}\right)
\ee
which leads to the following flow equations
\be
\left(\matrix{d\bff_t(s,t,\k)/d\ln\k&d\bfp_t(s,t,\k)/d\ln\k\cr
                     d\bpf_t(s,t,\k)/d\ln\k&d\bpp_t(s,t,\k)/d\ln\k\cr}\right)
\hspace{5.7cm}\nn
\ee
\be
\hspace{3.5cm} =\;\:\left(\matrix{-g_B^2K_m(s)&-g_B^2K_M(s)\cr
                    -g_B^2K_m(s)&0\cr}\right)
\left(\matrix{\bff_t(s,t,\k)&\bfp_t(s,t,\k)\cr
                     \bpf_t(s,t,\k)&\bpp_t(s,t,\k)\cr}\right)
\ee
It is straightforward to solve these equations using as initial conditions the
normalization conditions (12). One finds
\begin{eqnarray}
\lefteqn{\bff_t(s,t,\k)} \hspace{0.5cm}\nn \\
&=&{g_B^2\o2\k}\left(1+{(1+2{K_M(s)\o K_m(s)})\o
(1+4{K_M(s)\o K_m(s)})^{1\o2}}\right)
\left(t\o\k\right)^{\a_1(s)} + {g_B^2\o2\k}\left(1-{(1+2{K_M(s)\o K_m(s)})\o
(1+4{K_M(s)\o K_m(s)})^{1\o2}}\right)\left(t\o\k\right)^{\a_2(s)} \:,
\label{comp}\end{eqnarray}
\begin{eqnarray}
\lefteqn{\bpf_t(s,t,\k)=\bfp_t(s,t,\k)} \hspace{0.5cm}\nn \\
&=&{g_B^2\o2\k}\left(1+{1\o(1+4{K_M(s)\o K_m(s)})^{1\o2}}\right)
\left(t\o\k\right)^{\a_1(s)} + {g_B^2\o2\k}\left(1-{1\o(1+4{K_M(s)\o
K_m(s)})^{1\o2}}\right)
\left(t\o\k\right)^{\a_2(s)} \:,
\end{eqnarray}
\be
\bpp_t(s,t,\k)={g_B^2\o\k(1+4{K_M(s)\o K_m(s)})^{1\o2}}
\left(t\o\k\right)^{\a_1(s)} - {g_B^2\o\k(1+4{K_M(s)\o K_m(s)})^{1\o2}}
\left(t\o\k\right)^{\a_2(s)} \:,
\ee
where the Regge trajectories $\a_1$ and $\a_2$ are
\be
\a_1={g_B^2K_m(s)\o2}\left(1+\left(1+4{K_M(s)\o K_m(s)}\right)^{1\o2}\right)-1
\:,
\ee
\be
\a_2={g_B^2K_m(s)\o2}\left(1-\left(1+4{K_M(s)\o K_m(s)}\right)^{1\o2}\right)-1
\:.
\ee
It can be shown that these solutions correspond to the summation of leading
logs.
An entirely analogous consideration yields the functions $B_u^{\s
ijkl}(s,u,\k)$ which turn
out to be the same as the above with the simple change $t\ra -t$. The functions
$B_{B,s}^{\s ijkl}(t,u)$ on the other hand do not need renormalization
as their crossover is completely controllable within perturbation theory.

Finally then we find the two-particle--two-particle
$S$-matrix elements at one loop in the large $t$ limit to be
\begin{eqnarray}
\gt^{\ss \f\f\f\f}(s,t,\k)&=&{g_B^2 \o s + M^2} + {g_B^2\o 2\k}
\left(1+{(1+2{K_M(s)\o K_m(s)})\o
(1+4{K_M(s)\o K_m(s)})^{1\o2}}\right)\left\{\left(t\o\k\right)^{\a_1(s)}+
\left(-{t\o\k}\right)^{\a_1(s)}\right\} \nn \\
&& {}+ {g_B^2\o 2\k}\left(1-{(1+2{K_M(s)\o K_m(s)})\o
(1+4{K_M(s)\o K_m(s)})^{1\o2}}\right)\left\{\left(t\o\k\right)^{\a_2(s)}+
\left(-{t\o\k}\right)^{\a_2(s)}\right\} \:,
\end{eqnarray}
\begin{eqnarray}
\lefteqn{\gt^{\ss \f\f\p\p}(s,t,\k)\;\:=\;\:\gt^{\ss \p\p\f\f}(s,t,\k)\;\:=\;\:
{g_B^2\o2\k}\left(1+{1\o(1+4{K_M(s)\o K_m(s)})^{1\o2}}\right)
\left\{\left(t\o\k\right)^{\a_1(s)}+\left(-{t\o\k}\right)^{\a_1(s)}\right\}}
\hspace{4cm} \nn \\
&& {} + {g_B^2\o2\k}\left(1-{1\o(1+4{K_M(s)\o K_m(s)})^{1\o2}}\right)
\left\{\left(t\o\k\right)^{\a_2(s)}+\left(-{t\o\k}\right)^{\a_2(s)}\right\} \:,
\end{eqnarray}
\begin{eqnarray}
\gt^{\ss \p\p\p\p}(s,t,\k)&=&{g_B^2\o\k(1+4{K_M(s)\o K_m(s)})^{1\o2}}
\left\{\left(t\o\k\right)^{\a_1(s)}+\left(-{t\o\k}\right)^{\a_1(s)}\right\} \nn
\\
&& {} - {g_B^2\o\k(1+4{K_M(s)\o K_m(s)})^{1\o2}}
\left\{\left(t\o\k\right)^{\a_2(s)}
+\left(-{t\o\k}\right)^{\a_2(s)}\right\} \:.
\end{eqnarray}

Before we make our conclusions we will present the full
crossover function $\bff_t(s,t,\k)$ at one loop. For the sake of
simplicity, we consider here a theory
$(g_B/3!)\phi^3$, leaving a fuller discussion and extension to other cubic
theories for another publication. As mentioned above, apart from a
renormalization to get rid of the mass divergence, there is no need to have a
$t$ dependent renormalization of the coupling or the mass. However, in order
to arrive at a compact form for the crossover function here we will
implement the normalization conditions
\be
\Gamma^{\ss\f\f}(p^2=\k,\k)=\k+m^2(\k) \:,
\ee
\be
\Gamma^{\ss\f\f\f}(t=\k,\k)=g(\k) \:,
\ee
\be
\bff_t(s, t=\k, \k) = \frac{g^2(\k)}{\k + m^2(\k)} \:.
\ee
One finds for $\bff_t$
\be
\bff_t(s,t,\k)={g^2(t)\o t+m^2(t)}\exp\int^t_{\k} {d\k' \o \k'} \,
g^2(\k') f_1(\k') \:, \label{cross}
\ee
where
\be
g^2(t)=\frac{g^2(\k)}{1 - 2 g^2(\k)\left(f_2(t)-f_2(\k)\right)} \:,
\ee
\be
m^2(t)=m^2(\k)+\int^t_{\k} {d\k' \o \k'} \, g^2(\k') f_3(\k') \:.
\ee
The function $f_2$ is just the four dimensional one loop coupling correction,
whilst the functions $f_1$ and $f_3$ are given by
\be
f_1(\k)=\k{\partial\o\partial \k} \left( (\k +m^2(\k))h_1(\k) \right) \:,
\qquad
f_3(\k)=\k{\partial\o\partial \k}h_3(\k) \:,
\ee
$h_1$ being the one loop box diagram, and $h_3$ the one loop
mass correction. The function (\ref{cross}) crosses over to the
expression corresponding to (\ref{comp}) in the asymptotic limit
$t\gg \k\gg m^2$. In the small $t$ limit, if $g(\k)\ll1$,
then (\ref{cross}) just yields the perturbative one loop
expression for $\bff_t$ for the theory renormalized according to the
above normalization conditions.

To summarize, we have presented in this letter an
environmentally friendly renormalization
that is capable of capturing the crossover between large and small $t$
in the two-particle--two-particle scattering amplitudes. We derived in
particular the asymptotic behaviour in the case of a theory $\phi^2\psi$ with
unequal masses, including the relevant Regge trajectories and
derived a crossover scaling function for a $\phi^3$ theory. We hope that
it is clear that the methodology is capable of producing a wide variety of
results for the kinematical dimensional reduction that occurs in the Regge
limit.
In future publications we will present applications to more realistic theories
and
also implement an RG that is capable of accessing dual amplitudes.

\section*{Acknowledgements}

A.W. would like to thank the Instituto de Ciencias Nucleares for kind
hospitality.

\begin{figure}[p]
\centerline{
\hbox{\psfig{figure=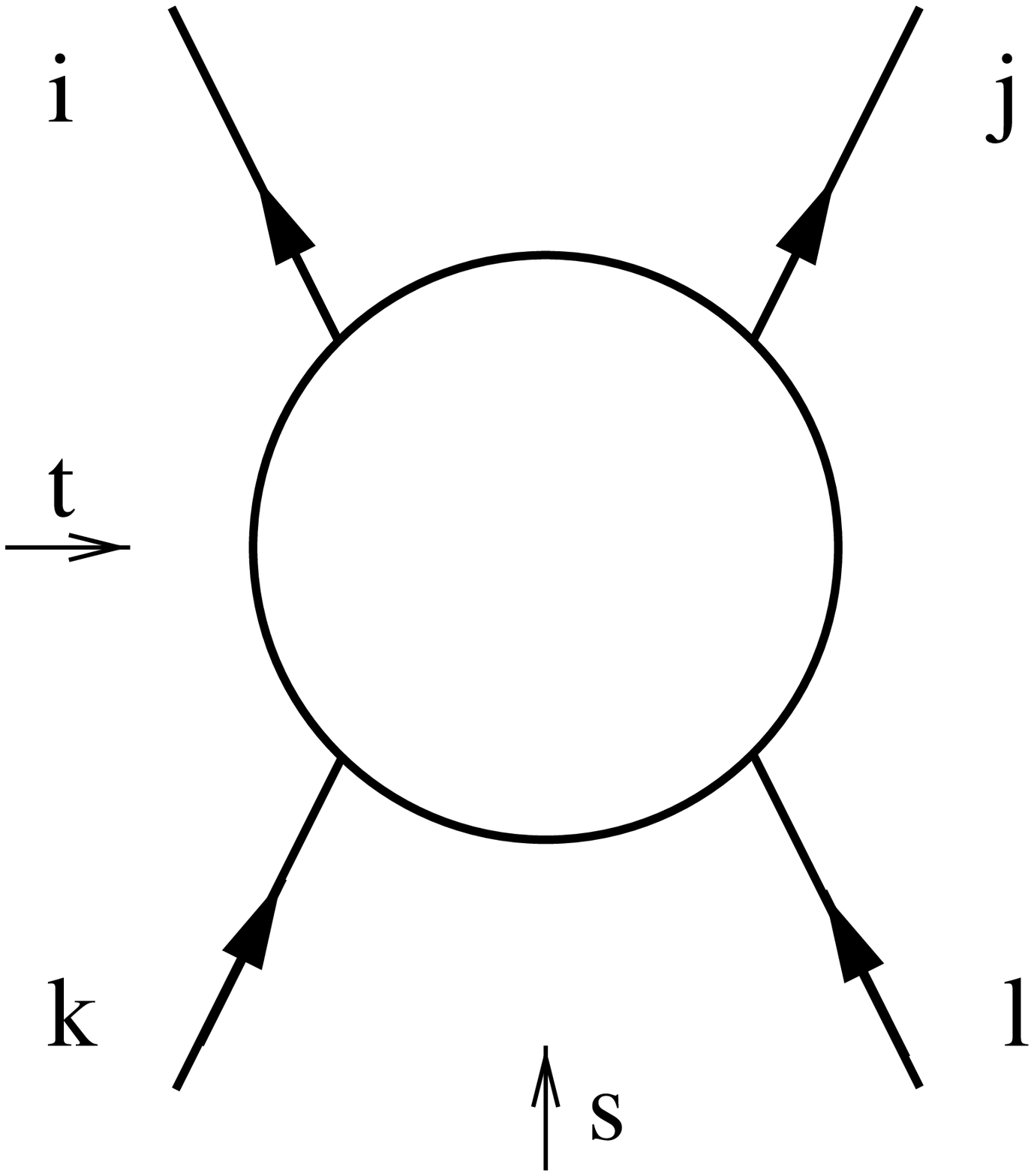,width=2in}}
\hspace*{0.2in}}
\par
\begin{center}
\bf Figure 1
\end{center}
\end{figure}


\begin{thebibliography}{99}
\bibitem{lipatov} E.A. Kuraev, L.N. Lipatov and V.S. Fadin,
{\it Sov.\ Phys.\ JETP\/} {\bf 45} (1977) 199.
\bibitem{eden} {\it ``The Analytic S-matrix''\/}, R.J. Eden, P.V. Landshoff,
D.I. Olive
and J.C. Polkinghorne; (CUP 1966).
\bibitem{migdal} A.A. Migdal, A.M. Polyakov and K.A. Ter-Martirosyan, {\it
Phys.\ Lett.\/}\
{\bf B48}
(1974) 239.
\bibitem{envfri} D.\ O'Connor and C.R.\ Stephens,
  {\it Nucl.\ Phys.}\ {\bf B360} (1991) 297;  {\it Int.\ J.\ Mod.\ Phys.}\ {\bf
A9} (1994)
2805;
  {\it Phys.\ Rev.\ Lett.}\ {\bf 72} (1994) 506.
\bibitem{fintemp} D.\ O'Connor, C.R.\ Stephens and F.\ Freire,
  {\it Mod.\ Phys.\ Lett.}\ {\bf A8} (1993) 1779; \\
M.A.\ van Eijck, C.R.\ Stephens and C.G.\ van Weert,
  {\it Mod.\ Phys.\ Lett.}\ {\bf A9} (1994) 309.

\end{thebibliography}
\end{document}